\documentclass[12pt]{article}
\begin{document}

\setlength{\baselineskip}{12.0pt}

\begin{center}
{\bf Curvature induced toroidal bound states} \\ {\vskip 6pt} Mario Encinosa and Lonnie Mott\\
Department of Physics \\   Florida A \& M University \\
  Tallahassee, Florida 32307
  \\
\end{center}
\setlength{\baselineskip}{16.0pt}
\begin{abstract} \noindent
 Curvature induced bound  state ($E < 0$) eigenvalues and eigenfunctions
for a particle constrained to move on the surface of a torus are
calculated. A limit on the number of bound states a torus with minor
radius $a$ and major radius $R$ can support is obtained. A condition for
mapping constrained particle wave functions on the torus into  free
particle wave functions is established. \vskip 6pt \noindent Pacs
number(s): 03.65Ge, 68.65.-k
\end{abstract}

%\section{Introduction}
\noindent$\bf{1. \ Introduction}$. \vskip 6pt The physics of
nanostructures [1,2] and quantum waveguides [3-7] may make questions
concerning curved surfaces in quantum theory increasingly relevant to
device modelling. Many workers have investigated with varying levels of
formal machinery the existence of bound states of quantum systems on
curved strips, tubes and hypervolumes [8-12]. The common thread through
much of the work described in [8-12] is the existence of an attractive
potential that appears in the Schrodinger equation as a consequence of
constraining a particle from  higher to  lower dimensional manifolds. (In
the majority of work the dimensionality is reduced from three to two, but
see [11] for the generalization to other cases.) This potential, called
here the curvature potential $V_C$, has been shown sufficient to cause
bound states in  model systems. In this work $V_C$ for a particle
constrained to the surface of a torus is derived and bound state surface
toroidal wavefunctions (STWs) are  calculated.

This brief report is organized as follows: in section 2 the method by
which $V_C$ is derived is concisely described for a symmetric but
non-trivial geometry. The method is then applied to the torus. In section
3 a brief description of the procedure used to solve the Hamiltonian
found in section 2 is given and some low-lying bound state eigenvalues
and STWs shown. Conclusions appear in section 4.

 \vskip 6pt
 \noindent
$\bf 2.\  Derivation\  of \  V_C $. \vskip 6pt

In the interest of clarity the derivation of $V_C$ will be performed for
a cylindrically symmetric surface. The extension to the general case is
straightforward and the salient points still obtain.

Let   ${\bf e}_{\rho},{\bf e}_{\phi},{\bf e}_{z}$ be  standard
cylindrical coordinate system unit vectors. A cylindrically symmetric
surface may be described by the Monge form
$$
{\bf r}(\rho,\phi)=\rho \ {\bf e}_{\rho} + S(\rho)\ {\bf e}_{z}.
\eqno(1)
$$
 $S(\rho)$ gives the shape of the surface. Points near the
 surface $S(\rho)$ may be described by
$$
{\bf x}(\rho,\phi,q)={\bf r}(\rho,\phi)+q \ {\bf e}_{n} \eqno(2)
$$
\noindent with ${\bf e}_{n}$ everywhere normal to the surface. The metric
near the surface $S$ is
$$
ds^2=  Z^2 \bigg[ 1-{{qS_{\rho\rho}} \over {Z^3}} \bigg ]^2 \ d\rho^2 +
\rho^2 \bigg[ 1-{{qS_{\rho}} \over {\rho^2Z}} \bigg ]^2\ d\phi^2 +\ dq^2
\eqno(3)
$$
$$
\equiv Z^2 [ 1+q k_1  ]^2 \ d\rho^2 + \rho^2 [ 1+qk_2]^2\ d\phi^2 +\ dq^2
\eqno(4)
$$
with subscripts indicating differentiation and $Z=\sqrt {1+S_\rho^2}$.
The Laplacian can be found straightforwardly from
$$
\nabla^2= g^{-{1 \over 2}}{\partial \over \partial q^i} \bigg [
g^{1 \over 2}\ g^{ij}{\partial \over \partial q^j} \bigg ],
\eqno(5)
$$
 but there is no advantage to writing the
the Laplacian explicitly until the constraint that places the particle on
the surface is effected.

 Consider a situation where a large confining potential everywhere normal to
$S$ acts to restrict the particle to $S$. This potential, called $V_n(q)
$ here, could take a hard wall or oscillator form, but however chosen it
causes $q \rightarrow 0$. In this limit the wave function is expected to
decouple into surface and normal parts, or in the language of [13], into
$\lq\lq$fast" and $\lq\lq$slow" functions
$$
\Psi(\rho,\phi,q) \rightarrow \chi_s(\rho,\phi)\chi_n(q). \eqno(6)
$$
Conservation of the norm in the decoupled limit implies [11,14,15]
$$
|\Psi|^2 WdSdq=|\chi_s|^2|\chi_n|^2dSdq \eqno(7)
$$
 where $W=1+2qH+q^2K$ and $dS$ the surface measure.  $H,K$ are the mean and Gaussian curvatures
given by
$$
H={1\over 2}(k_1+k_2), \eqno(8)
$$
$$
K=k_1k_2. \eqno(9)
$$
Write
$$\Psi = {\chi_s\chi_n \over \sqrt W}
\eqno(10)
$$
and insert the right hand side of equation (10) into the time
independent Schrodinger equation. Performing the differentiations
and taking $q\rightarrow 0$ gives the pair of equations
$$
-{1\over 2}\bigg[
 {1\over Z^2}{\partial^2 \over \partial \rho^2}
 +{1\over Z\rho}{\partial \over \partial \rho}
 -{Z_{\rho}\over Z^3}{\partial  \over \partial \rho}
 +{1\over \rho^2}{\partial^2 \over \partial \phi^2}
 +(H^2-K)\bigg]\chi_s  =E_s
\chi_s \eqno(11)
$$
$$
-{1 \over 2}{\partial^2  \chi_n \over \partial q^2} + V_n(q)
\chi_n =E_n \chi_n. \eqno(12)
$$
 The curvature potential $V_C$ is
$$
V_C = -{1 \over 2}[H^2-K]. \eqno(13)
$$

 The surface given by equation (1) was chosen to illustrate that
 there are modifications to the Laplacian (aside from the
 appearance of $V_C$) even for surfaces possessing symmetry.
 In [16] it was shown that kinetic energy modifications for
 some parameterizations of $S(\rho)$ can be small, but it
is easy to conceive of cases where the contrary would be true.

Now apply the above procedure to the torus. Let $F = R+a\ \rm cos
\theta$. Points near the surface of the torus may be parameterized as
$$
{\bf x}(\theta,\phi,q)=F{\bf e}_{\rho}+a\ {\rm sin}  \theta\ {\bf
e}_{z}+q \ {\bf e}_n. \eqno(14)
$$
Proceeding as above gives the constraint and curvature modified
Hamiltonian
$$
H = -{1 \over 2} \bigg [ {1 \over a^2}{\partial^2 \over \partial
\theta^2} -
 {{\rm sin}\ \theta \over aF}{\partial \over \partial \theta}+
{1 \over F^2}{\partial^2 \over {\partial \phi^2}} + {R^2\over a^2}
{1\over 4 F^2}
 \bigg ]
\eqno(15)
$$
$$
= H_0-{R^2\over a^2} {1\over 8 F^2}\equiv H_0+V_C.\eqno(16)
$$
In equation (16), $H_0$ is identical to $-{1\over 2}\nabla^2$ derived
  from eq. (14) with $q=0$; it is the Hamiltonian for  particle on a
 toroidal surface subject to no other potential [17] (it proves convenient to
 refer to the zero potential case as the free system).
 In contrast to the operator that appears in
equation (11), no surface dependent prefactors, i.e. $Z(\rho)$ or
additional terms modifying $H_0$ are present. It is interesting to
compare equation (15) to the corresponding operator for spherical and
cylindrical surfaces; for those surfaces the constraint procedure gives
a coordinate independent $V_C$
 behaving as $\sim -{1/R^2}$ [18] and again no
modifications to $H_0$. This point will be discussed further in
section 4. \vskip 6pt
%\section{Method}
 \noindent $\bf{3. \ Solution \ method;
\ results}$. \vskip 6pt \noindent Setting $\alpha = {a \over R},$ $\ $
$\beta = 2Ea^2$ and making the standard ansatz for the azimuthal
eigenfunction  $\chi(\phi) = exp \ [im\phi]$ in equation (15) gives
$$
  {\partial^2 \psi \over \partial \theta^2} -
 {\alpha \ {\rm sin}\ \theta \over [1 + \alpha \ \cos\theta]}{\partial \psi \over \partial \theta}
-{(m^2 \alpha^2- {1\over4}) \over [1 + \alpha \ \cos\theta]^2}\psi
+\beta\psi = 0. \eqno(17)
$$
Equation (17) can be solved numerically, but it is convenient to have
approximate analytic representations of its eigenfunctions. Recently a
method was found for obtaining closed form solutions for zero energy
states and for surface potentials $V_S(\theta)$ which satisfy an
auxiliary condition derivable from $H_0$ [19]. However, the
$V_S(\theta)=0$ case is no more easily solved with the method given in
[19] than the method employed below.

Solutions of equation (17) can be found by defining $z=exp[i\theta]$ and
writing
$$
\psi (z)= \sum^{\infty}_{n=-\infty}c_n\ z^n. \eqno(18)
$$

The Hamiltonian given by equation (15) is invariant under $\theta
\rightarrow  -\theta$, so the solutions of equation (17) can be split
into odd and even parity eigenfunctions, yielding a series in sines for
negative parity states  and cosines for positive parity states. Computing
the eigenvalues and eigenfunctions of equation (17) follows from a method
given in detail in [17].

Table 1 shows eigenvalues and wave functions for $\alpha = 3/4, 1/2, 1/4,
1/20$ for those $m$ values which yield $\beta<0$ states (only three
states are given for $\alpha = 1/20$; there are nine total). No negative
parity states appear in table 1.  In table 2
 wave functions for states corresponding to those in table 1
 with $V_C$ shut off are given.
As evidenced in table 2, the ground state $m=0$ wave function  of $H_0$
for any $\alpha$ is a constant.  A natural question is: Should the $V_C
\neq 0$ ground state wave function
 be compared to the constant wave function or
the lowest $\beta \neq 0$ wave function? Here the constant wave function
was chosen on the grounds that it is the state  actually  altered from
its constant value by $V_C$.

 \vskip 6pt
\noindent $\bf{4. \ Conclusions}$. \vskip 6pt
 In this brief report
wave functions for bound states of a particle constrained to the surface
of a torus were obtained for several values of $\alpha = a/R$. Constraint
and curvature effects were shown to alter the angular dependence of the
free particle STWs.

 An  interesting consequence of equation (17)
is manifested by the results presented in  table 1. For $m \neq 0$,
$\alpha = {1 \over 2m} $ provides a cutoff for the existence of bound
states. It follows that there are no $m \neq 0$ bound states for $\alpha
> 1/2$. Additionally $\alpha = 1/2m$  provides a series of magic radii
for which an $m\neq 0$ state of the constrained system maps exactly into
the $m=0$ state with the same $n$ and parity of the $V_C=0$ free STW.

In section 2 it was stated that the torus shares with the sphere the
property that constraint  adds only a curvature potential to the
Hamiltonian, leaving $\nabla^2$ on the surface unchanged.  This
comparable behavior is likely a consequence of  the torus being the most
symmetric compact genus one surface that can be embedded in $R^3$. It
would be interesting to learn if there are higher genus surfaces embedded
in $R^3$ for which $H=-{1\over 2} \nabla^2$ reduces to the lower
dimensional operator $H_0$ plus a curvature potential upon imposing the
condition given by equation (10). This question generalizes to $M<N$
dimensional surfaces embedded in $R^N$. \vskip 6pt

\vfill \eject \centerline{\bf Acknowledgments} \vskip 6pt \noindent The
authors would like to acknowledge useful discussions with Babak Etemadi.
M.E. would like to thank Norman Melom for useful suggestions. Both
authors received support from NASA grant NAG2-1439. \vskip 6pt
% \vfill\eject \vskip 6pt

\centerline{\bf References}
 \vskip 6pt
 \noindent
 1. S. M. Reimann and M. Manninen, Rev. Mod. Phys. ${\bf 74}$, 1953
 (2002).
 \vskip 6pt
\noindent
2. D. Ferry and S. Goodnick, ${\it Transport \ in \
nanostructures}$, (Cambridge University Press, U.K., 1992).
 \vskip 6pt
\noindent
 3. P. Ouyang, V. Mohta and R. L. Jaffe, Ann. of Phys. ${\bf 275}$,
297 (1998).
  \vskip 6pt
\noindent
 4. I. Y. Popov, Phys. Lett. A ${\bf 269}$, 148 (2000).
\vskip 6pt
 \noindent
 5. S. Midgley, Aus. J. Phys ${\bf 53}$, 77 (2000).
 \vskip 6pt
 \noindent
6. J.T. Londergan, J.P. Carini, D.P. Murdock, ${\it Binding \ and \
scattering  \ in \ two \ dimensional}$
\noindent
 ${\it \ systems;\ applications \ to \
 quantum \ wires, \ waveguides, \ and \ photonic \ crystals}$,
(Springer-Verlag Berlin, 1999). \vskip 6pt \noindent
 7. J. Goldstone and R. L. Jaffe, Phys. Rev. B ${\bf 45}$, 14102 (1991).
 \vskip 6pt
 \noindent
 8. I. J. Clark and A. J. Bracken, J. Phys. A ${\bf 29}$, 4527 (1996).
 \vskip 6pt
 \noindent
 9. P. Duclos and P. Exner, ${\it
Reviews \ in \ mathematical \ physics}$, ${\bf 7}$, 73 (1995). \vskip 6pt
\noindent
10. D. Krejcirik, J. Geom. Phys. ${\bf 45}$, 203 (2003). \vskip
6pt \noindent
11. P. C. Schuster and R. L. Jaffe, hep-th /0302216
\vskip
6pt \noindent
12. R. C. T. da Costa, Phys. Rev. A ${\bf 23}$, 1982
(1981).
  \vskip 6pt
  \noindent
13. L.Kaplan, N.T. Maitra and E.J. Heller, Phys. Rev. A ${\bf 56}$,
 2592 (1992).
\vskip 6pt \noindent 14. R. C. T. da Costa, Phys. Rev. A ${\bf 25}$, 2893
(1982). \vskip 6pt \noindent
 15. M. Encinosa and B. Etemadi, Physica B ${\bf 266}$, 361 (1998).
 \vskip 6pt \noindent
16. M. Encinosa and B. Etemadi, Phys. Rev. A ${\bf 58}$, 77 (1998).
\vskip 6pt \noindent
 17. M. Encinosa and B. Etemadi, quant-ph /0200501, and submitted to Found.
 Phys. Lett.
 \vskip 6pt
 \noindent
 18. H. Jensen and H. Koppe, Ann. of Phys. ${\bf 63}$, 586 (1971).
 \vskip 6pt
\noindent 19. A. Schulze-Halberg, Found. Phys. Lett. ${\bf 15}$, 585
(2002).

\vfill \eject
 \addtolength{\topmargin}{-0.5in}
\begin{table}
\caption{Eigenvalues and wave functions of equation (17) for four values
of $\alpha$. Normalizations (not including the $(2\pi)^{-{1\over2}}$ from
the $\phi$ dependence) are in brackets proceeding the functions. Terms
not shown are at least an order of magnitude smaller than those listed.
There are no negative parity states with $\beta <0$.}
\begin{center}
\begin{tabular}{|l|l|l|}
\hline
\ \    $\alpha$ & $\quad \ \beta$ &  $\Psi_{nm}(\theta); \ \  V_C \neq 0$ \\
\hline
.75 & -1.0725 & $\Psi_{10}=(.1298)[4.6072-\ 5.2143 \ \rm cos \theta+2.2465\ \rm cos2 \theta-.9495\ \rm cos3 \theta]$\\
.50 & -0.3512 & $\Psi_{10}=(.2455)[2.4509-\  .9015\ \rm cos \theta+.1921\ \rm cos 2 \theta]$  \\
.25 & -0.2673 & $\Psi_{10}=(.3765)[2.1458-\ .2916\rm cos\theta+\ .0280\rm cos2\theta]$  \\
.25 & -0.1987 & $\Psi_{11}=(.3826)[2.1069-\ .2138\rm cos\theta+\ .0197\rm cos 2\theta]$  \\
.05 & -0.2506 & $\Psi_{10}=(.8813)[2.0254-\ .0508\rm cos\theta]$  \\
.05 & -0.2481 & $\Psi_{11}=(.8814)[2.0251-\ .0507\rm cos\theta]$  \\
.05 & -0.2406 & $\Psi_{12}=(.8817)[2.0244-\ .0487\rm cos\theta]$  \\
 \hline
\end{tabular}
\end{center}
\end{table}

\begin{table}
 \caption{ Eigenvalues and wave functions of $H_0$ corresponding to
those appearing in table 1.}
\begin{center}
\begin{tabular}{|l|l|l|}
\hline
\ \ $\alpha$ & $\quad \ \beta$ &  $\Psi_{nm}(\theta); \ \  V_C = 0$ \\
\hline
.75 & 0.0000 & $\Psi_{00}=.4607$\\
.50 & 0.0000 & $\Psi_{00}=.5642$  \\
.25 & 0.0000 & $\Psi_{00}=.7979$  \\
.25 & 0.0641 & $\Psi_{11}=(.4073)[1.9676+\ .0648\rm cos\theta]$  \\
.05 & 0.0000 & $\Psi_{00}=1.7841$  \\
.05 & 0.0025 & $\Psi_{11}=(.8822)[1.9998+\ .0005\rm cos\theta]$  \\
.05 & 0.0010 & $\Psi_{12}=(.8822)[1.9996+\ .0002\rm cos\theta]$  \\
 \hline
\end{tabular}
\end{center}
\end{table}
\end{document}